\documentclass[journal]{IEEEtran}
\ifCLASSINFOpdf \else \fi
\usepackage{url}
\usepackage{color}
\usepackage{color,array}
\usepackage{lipsum}
\usepackage{hhline}
\usepackage[table,xcdraw]{xcolor}
\usepackage{amssymb,amsmath,color,graphicx, amsbsy}
\usepackage{graphicx}
\usepackage{subfigure}
\usepackage{cite}
\usepackage[linesnumbered,ruled]{algorithm2e}
\usepackage{amsthm}
\usepackage{amsfonts}
\usepackage{amssymb}
\usepackage{graphicx}
\usepackage{epstopdf}
\usepackage{makeidx}
\usepackage{outlines}
\usepackage[noprefix]{nomencl}
\usepackage{textcomp}
\usepackage{soul}
\usepackage{multirow}
\usepackage{hyperref}

\hypersetup{
	colorlinks=true,
	linkcolor=black,
	filecolor=black,
	urlcolor=black,
	anchorcolor	=black,
	citecolor=black,
}
\allowdisplaybreaks
\usepackage{graphicx}

\begin{document}

\title{Autonomous Charging of Electric Vehicle Fleets to Enhance Renewable Generation Dispatchability}


\author{ {Reza Bayani, \emph{Student Member, IEEE} and Saeed D. Manshadi, \emph{Member,  IEEE}, Guangyi Liu, \emph{Senior Member, IEEE}, Yawei Wang, \emph{Student Member, IEEE},  Renchang Dai,  \emph{Senior Member, IEEE}}
\thanks{R. Bayani and S. D. Manshadi are with San Diego State University, San Diego,
CA, 92182 USA. e-mail: rbayani@sdsu.edu; smanshadi@sdsu.edu. G. Liu, Y. Wang, and R. Dai are with GEIRI North America, San Jose, CA, 95134. e-mail: guangyi.liu@geirina.net; renchang.dai@geirina.net; yawei@gwmail.gwu.edu.
This work is supported by State Grid Corporation technology project 5100-201958522A-0-0-00}
}


\maketitle

\begin{abstract}
A total 19\% of generation capacity in California is offered by PV units and over some months, more than 10\% of this energy is curtailed. In this research, a novel approach to reduce renewable generation curtailments and increasing system flexibility by means of electric vehicles' charging coordination is represented. The presented problem is a sequential decision making process, and is solved by fitted $Q$-iteration algorithm which unlike other reinforcement learning methods, needs fewer episodes of learning. Three case studies are presented to validate the effectiveness of the proposed approach. These cases include aggregator load following, ramp service and utilization of non-deterministic PV generation. The results suggest that through this framework, EVs successfully learn how to adjust their charging schedule in stochastic scenarios where their trip times, as well as solar power generation are unknown beforehand. 
\end{abstract}

 
 \section*{Nomenclature}
 \subsection*{Variables}
\noindent \begin{tabular}{ l p{7.25cm} } 

$a_t$ & Action at time $t$\\
$p_t^{bought}$ & Total amount of energy bought from electricity grid at time $t$ by EV fleet\\
$p_{(i,t)}$ & Charging power of the EV indexed by $i$ at time $t$\\
$r_t$ & Reward of taking an action at time $t$\\

\end{tabular}
\subsection*{Parameters}
\noindent \begin{tabular}{ l p{6cm} }
$D$ & Total number of simulation days\\
$K_{max}$ & Maximum number of iterations\\
$PV_t$ & PV output at time $t$ (kW)\\
$\mathcal{\mu}_{D}, \mathcal{\mu}_{A}$ & Mean departure and arrival time for vehicles of EV fleet (hr)\\
$\mathcal{\mu}_i^d, \mathcal{\mu}_i^a$ & Mean departure and arrival time for single EV indexed by $i$ (hr)\\
$\mathcal{\sigma}_{D}, \mathcal{\sigma}_{A}$ & Variance of departure and arrival time for vehicles of EV fleet (hr)\\
$\mathcal{\sigma}_i^d, \mathcal{\sigma}_i^a$ & Variance of departure and arrival time for single EV indexed by $i$ (hr)\\

\end{tabular}
\subsection*{Sets}
\noindent \begin{tabular}{ l p{6cm} }
$\mathcal{A}$ & Set of actions\\
$\mathcal{D}$ & Set of days\\
$\mathcal{I}$ & Set of all EVs\\
$\mathcal{S}$ & Set of states\\
$\mathcal{T}$ & Set of time steps\\

\end{tabular}

\section{Introduction} 
\IEEEPARstart{D}UE to their environment friendly nature and unlimited sources of supply (e.g. sunlight and wind), renewable generation units are increasingly being integrated into the power system. Based on the statistics provided by  U.S. Energy Information Administration, currently, renewable resources account for 19\% of total electricity generations of the United States and are expected to double their share among generation fleet (i.e. natural gas, renewable, coal and nuclear) in 30 years, rendering them the major sources of energy in United States by 2050 \cite{eia2020}. Among renewable generations, currently wind turbines are the pioneer type, being responsible for 38\% percent of renewable generation fleet in the United States. Solar generations currently provide 15\% of renewable generation output, but it is projected that photovoltaic (PV) resources will contribute 46\% of total renewable generation by 2050, which means a considerable 18\% of total electricity generation of U.S. will be provided solely by PV units. California is the second largest consumer of electricity energy in the United States and in 2018, California ranked first in the nation as a producer of electricity from solar resources \cite{eia2020}. In 2018, large and small scale solar PV and solar thermal provided 19\% of California’s net electricity generation. 
\newline
Along with the several benefits of PV integration, system operators are facing new challenges risen as a result of introducing substantial amounts of PV, such as over-generation which leads to periods of large amounts of curtailment. Primarily, curtailment is executed to reach supply-demand balance and avoid over-voltage instances in the network\cite{kikusato2019electric}. Although an easily accessible solution, curtailing is a waste of resources which diminishes the investor confidence\cite{golden2015curtailment}. Also to reach a higher utilization of generation fleet, it is desirable to reduce PV generation curtailments.\newline
\begin{figure}[h!]
\centering
{\includegraphics[width=8.8cm]
{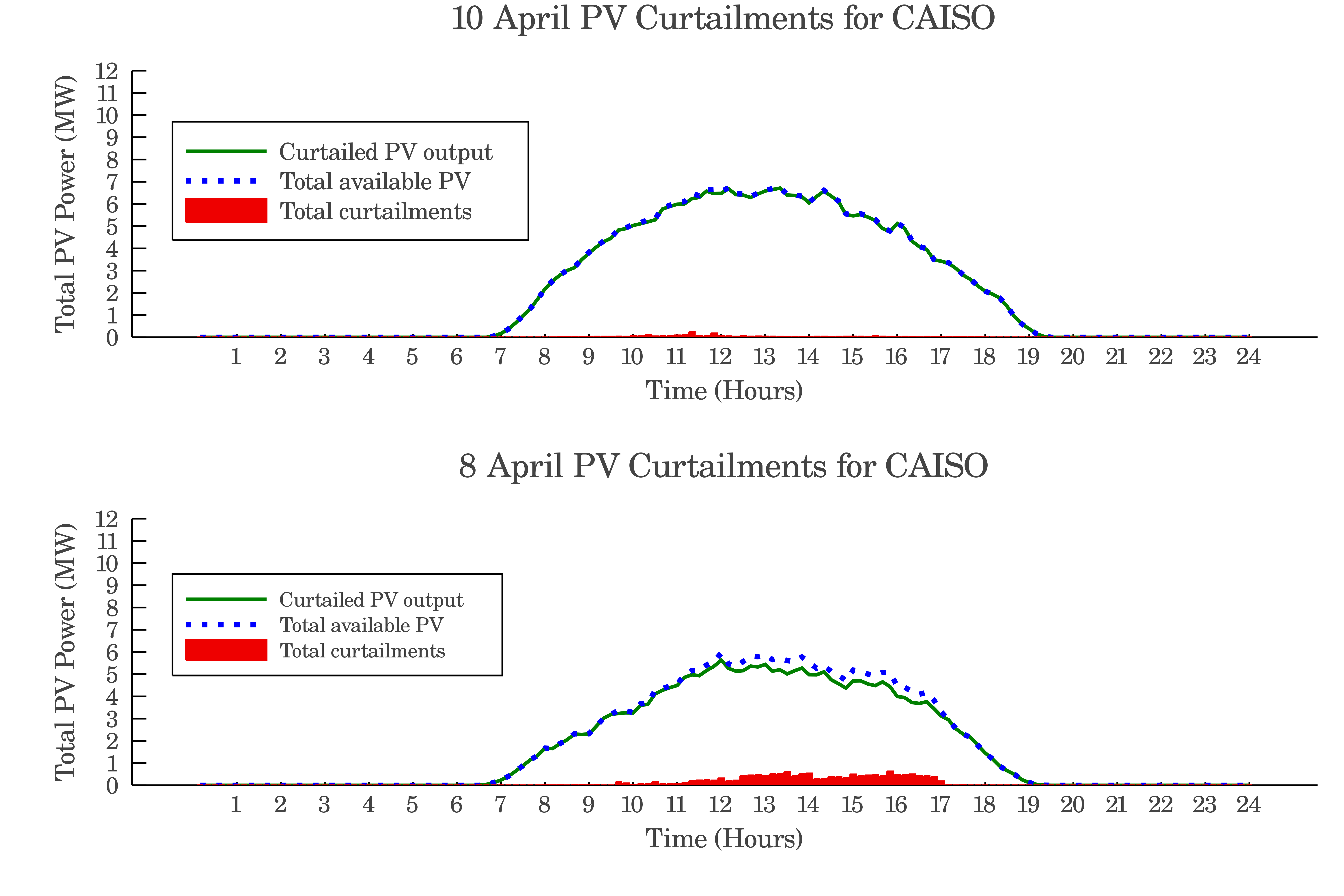}}
\end{figure}
\begin{figure}[h!]
\centering
{\includegraphics[width=8.6cm]
{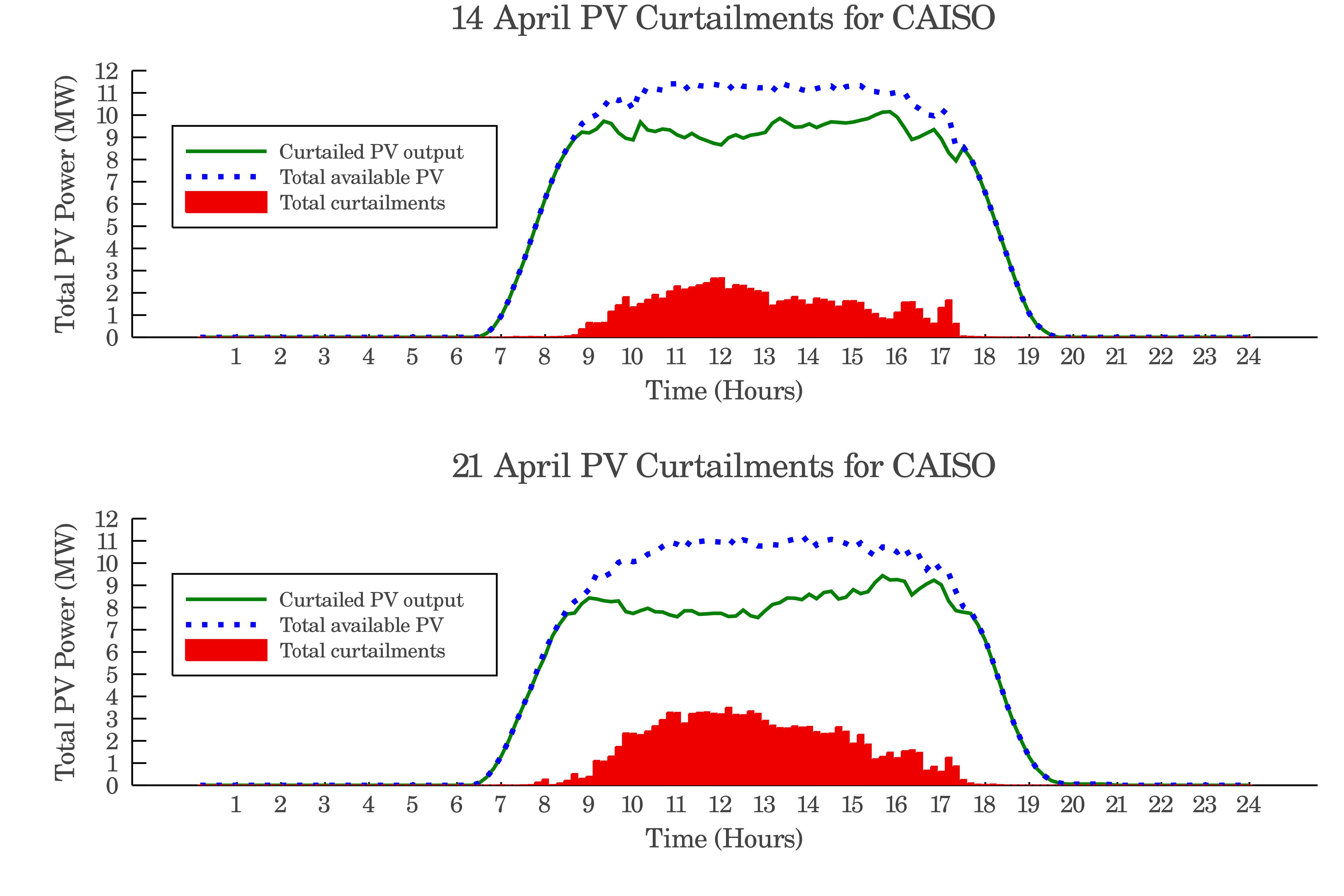}}
\caption{Comparison of California PV output before and after curtailments}
\label{fig:curtailment compared}
\end{figure}
According to the data provided by California Independent System Operator (CAISO), 10\% of the state's total PV generation has been curtailed in the first four months (January through April) of 2020 \cite{caisowebsite}. The PV output curves for a number of selected days in April of 2020 for CAISO are portrayed in Fig. \ref{fig:curtailment compared}. As can be seen from this figure, for $10^{th}$ of April, less than 1\% of PV output was curtailed. One reason is that for this day, the peak of supply did not exceed the peak of demand as the weather was not clear and PV output has not reached its full capacity. The amount of curtailment for $8^{th}$ of April exceeds 5\% and as the total PV output is increased, like for $14^{th}$ and $21^{st}$ of April, the curtailment is also increased dramatically, adding up to 12.2\% and 17.7\% for these days, respectively. Therefore, it is important to come up with ways to address curtailment reduction of the PV resources, which as highlighted earlier, are the fastest growing generation units of any type.

 The dramatic increase in the share of renewable energy will cause more power curtailments which negatively affects the value and cost of renewable generation units. Insufficient transmission, congestion, and excessive supply are counted as the major causes of curtailments\cite{cochran2015grid}. Several practical measures are being taken by the industry to reduce curtailments. In Colorado, Automatic Generation Control (AGC) is used to deal with the curtailments in wind generation \cite{cochran2012integrating}. To mitigate curtailments, a market approach which CAISO has implemented is negative pricing during oversupply periods \cite{bird2014wind}. Authors in \cite{li2015comprehensive} count enhancement of operational reliability, generation flexibility and maintenance plans of renewable energy generators as solutions to reduce curtailed electric energy from technological viewpoint. Another practical means of dealing with curtailments reduction is transmission expansion and augmenting interconnections, which is currently being practiced by ISO New England, Midcontinent ISO and PJM Interconnection \cite{bird2014wind}. Storage technologies \cite{golden2015curtailment}, and utilization of curtailed resources for ancillary services such as frequency regulation \cite{bird2014wind} are also counted as measures to reduce curtailments.\\ 
Improved forecasting and economic dispatch \cite{henriot2015economic,ela2009using}, and storage technologies \cite{zou2015mitigation,cleary2015assessing,moradzadeh2014congestion,denholm2019timescales,hozouri2014use} are among approaches used by researches to address the curtailment issue. To offer more operational flexibility by reducing curtailments in a microgrid with high penetration levels of renewable generation, the authors of \cite{denholm2019timescales} have investigated energy storage solutions. They show that compared to a case with no storage, adding storage devices with only 4 hours of daily operation can reduce the curtailments from 16\% to 10\%. By considering congestion and load mismatch as reasons for wind turbine curtailments, authors in \cite{hozouri2014use} have proposed a joint operational scheme for wind power generation system with pumped hydro energy storage. 
\\In this paper, we have introduced a management scheme based on Electric Vehicles (EV) cooperation to schedule the charging of them to reduce PV curtailments. EVs of any type have been experiencing a surge in demand due to reasons such as no pollutant productions, technological advancements leading to reductions in price and increased range, government policies encouraging customers, etc. It is estimated that the combined share of non-electric vehicles in the United States will fall from the current 94\% to 81\% by 2050\cite{eia2020}. If done properly, EV fleet management not only prevents the potential negative impacts of EV charging on the power grid operation, but also could be beneficial for both the owners and operators. \\
One of the challenges of problem settings containing EVs is the randomness of their behavior such as arrival and departure times, and reinforcement learning (RL) algorithms are capable of coping with these problems. RL is a promising tool for problems where the system model is large, complex and not determined and is increasingly being implemented in power system applications. In this branch of machine learning, the learning takes place based on the interactions between the environment and the agent. RL offers solutions for stochastic sequential decision making problems and various research is being conducted on RL applications in power systems\cite{zhang2018review}. The applications include but are not limited to energy management\cite{mocanu2018line, szinai2020reduced}, demand response\cite{chics2016reinforcement}, electricity market \cite{chen2018indirect}, and operational control\cite{claessens2016convolutional}. The common learning algorithms utilized in RL are $Q$-Learning\cite{watkins1992q}, fitted $Q$-Iteration\cite{ernst2005tree}, and Deep Reinforcement Learning (DRL) approaches \cite{mnih2015human, zhang2019deep}.\newline
Authors in \cite{da2019coordination} have presented a multi-agent framework to charge EV batteries with the objective of reducing transformer overloads. In this study, RL agents learn two policies, one is a selfish one that tends to increase local reward and the collaborative policy, which looks at other agents' preferences as well. In another study, the charging management problem of EVs has been addressed by a DRL algorithm so as to reduce the overall travel time and charging costs at electric vehicle charging stations\cite{qian2019deep}. To deal with the uncertainties, first a deterministic model is extracted which then is used to attain transition probabilities. The charging/discharging schedule and location of EVs are extracted in \cite{ko2018mobility} through a mobility aware control algorithm that takes EV mobility, SOC, and demand of the system into consideration to solve RL with the value iteration method. A DRL model is proposed in \cite{li2019constrained} to acquire the optimal charging/discharging schedules of EVs to ensure full charging of EVs while taking the stochasticity of electricity prices into account.\\
The unpredictable variables such as traffic conditions and electricity prices inherent to an EV scheduling problem are dealt with a model-free approach in \cite{wan2018model}. It is shown the desired solution can be obtained by adaptively learning the transition probabilities. A multi-agent DRL approach is proposed in \cite{shin2019cooperative} to handle various EV charging stations with varying dynamics. Another model for EV charging scheduling problem of a park-and-charge system which aims to minimize battery degradation costs can be found in \cite{wei2018electric}. The problem with these methods is their dependency to large scale data, and with not big enough experience samples these approaches will not succeed.

As discussed, a variety of problems in power system applications have been addressed with $Q$-learning and DRL methods. A major shortcoming of $Q$-learning is that for the settings where state and action spaces are not finite or small, the Q value can no longer be tabularized\cite{ernst2005tree}. On the other hand, although effective in many applications because of the intrinsic power of neural networks in approximating functions, DRL does not become much of help when the available dataset size is not very large. Thus, in a problem setting with limited sets of data and large/sparse state action tuples, fitted $Q$-iteration algorithm suits best.\newline
To reach the ideal day-ahead charging plan of an EV fleet, a heuristic plan is proposed in \cite{vandael2015reinforcement}. With the objective of reducing costs, the beforehand unknown charging flexibility of EVs, which is dependent on various factors such as plug-in times, power considerations, and battery specifications are effectively learned in this work by batch reinforcement learning.
A demand response model-free approach to jointly coordinate the charging stations in a network, a batch RL algorithm has been introduced in \cite{sadeghianpourhamami2019definition}. Instead of considering EVs as a whole, a novel method is presented here to control single EVs and it is shown that a suitable optimal policy compared to the benchmark all knowing oracle can be achieved by this method. The flexibility of supply and storage devices in a multi-carrier energy system has been investigated in \cite{mbuwir2017battery}. This sequential decision making process dependent on stochastic features such as weather and electricity demand has been tackled with a fitted $Q$-iteration algorithm. Results show that without fully knowing the model of the system, an optimal interaction of system carriers can be achieved.\newline
The energy management problem of a microgrid is studied with the batch Reinforcement Learning method in \cite{mbuwir2017battery}. With the objective of maximizing the usage of the connected PV device, the control policy of a storage unit is obtained through a data-driven fitted $Q$-iteration. In another study, a demand response approach for minimizing the charging costs of a single plug-in electric vehicle is proposed\cite{chics2016reinforcement}. Here, different charging scenarios are fed to the regression model as historical data with non-deterministic energy prices. The results support the effectiveness of batch RL algorithm in the reduction of the charging prices.\\
 In a nutshell, in the presented work, an EV fleet is utilized as a means of offering flexibility, rather than relying on conventional methods such as storage and market solutions. The main contribution of this work is to present a novel approach which reduces renewable generation curtailments by means of electric vehicles' autonomous charging coordination. Besides that, an RL framework is successfully implemented in the presence of uncertainties in PV generation such that the curtailment reductions are mitigated. Finally, this problem setting is also applied to a DRL algorithm and it is verified that with the fitted $Q$-iteration algorithm, Q-function can be learned over only a limited number of simulation days rather than prolonged training periods, which is the case in some other DRL algorithms.\\
 The problem of scheduling EV charging can be divided into two steps. First, the environment is modeled and then, fitted $Q$-iteration algorithm is implemented. Section II is designated to discuss the problem formulation by introducing some concepts in sequential decision making problems and RL; while in section III the solution methodology and algorithm are explained. The results for three case studies,  as well as performance comparison of our algorithm vs. a DRL algorithm are analyzed in section IV.  Eventually, in section V, the motive and findings of this research are recapitulated with some suggestions on future work.
\section{Problem Formulation}
In this section, the environment setting modeling as a finite Markov Decision Process (MDP) is explained. Later on, the formulations used in this work to represent and capture the problem are discussed in detail.
\subsection{Finite Markov Decision Processes}
\begin{figure}[h]
\centering
{\includegraphics[width=5cm]
{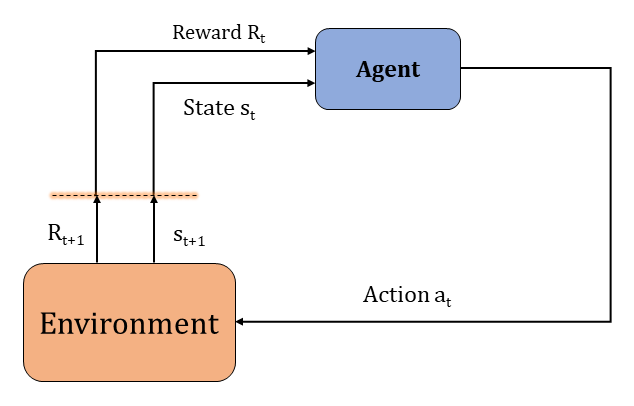}}
\caption{The agent-environment interaction in MDPs}
\label{fig:env agent}
\end{figure}

In sequential decision making problems, actions impact both the outcome and the next state of the system. MDPs are a formalized way to present these problems. In a finite MDP, rewards and states depend only on the previous state and action. MDPs are defined with state space $\mathcal{S}\subset\mathbf{R}^d$, action space $\mathcal{A}\subset\mathbf{R}$, transition probability $p(s_{t+1}\mid s_t,a_t)$ which defines the dynamics of the system, and reward function given as $R(s_t,a_t)$. Fig. \ref{fig:env agent} shows how the agent and the environment interact in an MDP setting.
\\
According to Fig. \ref{fig:env agent}, the agent at each state observes the current state and reward signals of the system, i.e. $s_t$ and $r_t$, respectively. The agent could be an aggregator who aims to match the demands of EVs it controls with its bought energy, or a PV owner who tries to maximize the utilization of PVs. The agents sends the action signals, $a_t$, to the environment, which could be the aggregate charging power of EV fleet or total PV generation in these examples, and wait for the environment's response to these actions at the next time step.
\subsection{Environment Setting}
As mentioned in the previous part, the actions are directed to the environment at each state, and the environment response is observed as the next state and reward. EV owners are a source of uncertainty in the system as their behavior is not deterministic. It is assumed that each EV owner departs home in the morning, reaches the charging station at the workplace, then heads back home in the evening and connects to the charger in the home to attain a desirable level of charging by the next morning. In order to achieve a better estimate of the EV fleet behavior, we could differentiate between types of EVs, different battery sizes, days of the week (workdays or weekends) or even seasons of the year. However, as the contribution of this work is not affected by any of these, we have considered same vehicle specifications for all of the fleet vehicles and have not distinguished daily patterns.\\
For each EV owner, mean of arriving and departure time are derived from the normal distribution $\mathcal{N}(\mu_{D}, \sigma_{D})$ and $\mathcal{N}(\mu_{A}, \sigma_{A})$, respectively. Next, for each day, the departure and arrival hours of each EV are drawn from the normal distributions $\mathcal{N}(\mu_i^d, \sigma_i^d)$ and $\mathcal{N}(\mu_i^a, \sigma_i^a)$, respectively. Also, in addition to the uncertainty of EV departure and arrival times, the SOC of each EV when reaching a destination is also an stochastic variable. 
\newline
The state of the environment at each time step is an n-tuple encapsulating the essential characteristics of the model often regarded as \textit{features} to reach the best learning experience. Feature selection is one of the most important aspects of problem solving and can impact accuracy as well as the complexity of the system. For this problem, if the states were to be defined according to the SOC and availability of single EVs, that would have led to a gigantic state space size. The same applies to actions, thus in order to reduce the state action space size, state elements have been introduced so they contain vital information signaling the overall state of the aggregated cars in EV fleet. The charging stations are presumed to be able to measure the SOC of connected EVs and with appropriate communication links, the state can be considered observable at all times. It is assumed that after connecting to the charging station at each instance, the drivers set the desired time for leave, so that information is also known. By knowing the battery specifications such as charging rate and capacity, the charging time required for each EV can be calculated at any time.\\
The set of all EVs in the system are represented with $\mathcal{I} = \left\{ 1,2,..,N \right\}$, while the set of all available EVs, $\mathcal{C}$, consists of EVs $\in$ $\mathcal{I}$ which are stationary, i.e. connected to the charging station. The set of all available EVs for charging and the set of all critical EVs, denoted respectively by $\mathcal{V}$ and $\mathcal{U}$, are defined in \eqref{set_v} and \eqref{eq:set critical}. 

\begin{equation}
    \mathcal{V} = \left\{ EV_i \right\}: i\in\mathcal{C} \And SOC_i<100 \label{set_v}\\ 
\end{equation}
\begin{equation} \label{eq:set critical}
    \mathcal{U} = \left\{ EV_i \right\}: i\in\mathcal{V} \And t_i^a-t_i^r\leq\delta_c \\
\end{equation}
\color{black}In \eqref{eq:set critical}, $t_i^a$ and $t_i^r$  are the available time window for charging of $EV_i$ and the required total time that is needed for $EV_i$ to reach full charge, respectively, and it is assumed they are known for all EVs that are connected to charging stations. The relation between $t_i^a$ and $t_i^r$ is not forced directly. However, for the EVs that are connected at home, the algorithm works in a way that $t_i^a$ is always greater than $t_i^r$. The reason is that according to \eqref{eq:set critical}, if the difference between $t_i^a$ and $t_i^r$ for a vehicle reaches $\delta_c$, that EV is moved to the set of critical EVs. This means this particular EV is charged by force at the next control step, and its $t_i^r$ is reduced. At every control step, EVs in $\mathcal{U}$ will be charged through ``forced actions", assuring all EVs will reach 100 \% SOC before desired time. It is also worthwhile to mention that the problem is defined in such a way that it is feasible for each EV to become fully charged during the night. EV technology has some limitations and the waiting time for full charging is on top of the list. The idea of ``in motion charging" of EVs is an attempt to circumvent this limitation, and is further discussed in our previous works \cite{manshadi2017wireless,manshadi2019strategic}. Wireless charging allows EV owners to leave before EVs are reached full charged status, as they can be charged even during the trip.\\
\color{black}$n_{max}$ and $n_{min}$ are the cardinality of $\mathcal{V}$ and $\mathcal{U}$ at each time step, respectively. $n_{max}$ sets the upper limit for the number of EVs that can be charged at a certain time step, while $n_{min}$ is equal to the minimum number of EVs that must be charged at a certain time. The set of feasible actions at each time, $\mathcal{H}_t$ is defined in \eqref{eq:feasbile actions}. During the learning process at every time step, the agent selects its next action from this set. In \cite{da2019coordination}, whether the EV is available for charging or not is considered in the state space. In another application for frequency control with EVs' batteries, the number of vehicles that have SOC levels below (above) minimum (maximum) acceptable SOC level are also presented in the state space \cite{wang2020v2g}. In our problem, the overall availability of the EVs is estimated by presenting $n_{min}$ and ${n_{max}}$ to the state tuple.\\ 
\begin{equation} \label{eq:feasbile actions}
    \mathcal{H}_t = \left\{\begin{matrix}
\left\{n_{min}:n_{max}\right\}, & n_{max}>0\\
&\\
0, & n_{max}=0
\end{matrix}\right. 
\end{equation}
The total amount of time needed for all EVs to be fully charged, $T_t$, is another constituent of the state and is acquired as shown in \eqref{eq:total time required}.  Authors in \cite{vandael2015reinforcement} have used the variable $E_{req}$ in the state tuple of the system. In our work, is assumed that all of the batteries have the same characteristics, and the EVs are charged (discharged) at same rates. This means that the total amount of energy which is needed for all of the EVs to be charged is proportional to the variable $T_t$. Hence, this variable will provide an accurate measure of the overall average SOC of the fleet. 
\begin{equation} \label{eq:total time required}
    T_t = \sum_{i\in\mathcal{I}}t_i^r
\end{equation}
\\Freedom is the most important variable in the presented problem formulation, as four features of the state are dependent on it. The freedom of each EV at each time is calculated as given in \eqref{eq:flex}.
\begin{equation} \label{eq:flex}
    Fr_{(i,t)} = \left\{\begin{matrix}
\frac{t_i^a-t_i^r}{t_i^a}, & i\in\mathcal{V}\\
&\\
1, & i\in\mathcal{I}\setminus\mathcal{C}
\end{matrix}\right. 
\end{equation}
According to \eqref{eq:flex}, the freedom of $EV_i$ is a number between $[0,1]$, and the lower the freedom of an EV is, the less margin is available for delaying its charging. We have concluded that placing four representations in states which are acquired from the freedom of EVs, i.e. $\left\{Fr_{ave},Fr_2,Fr_5,Fr_{10}\right\}$, will go a long way toward learning.  The freedom variable presented in this work is a measure of how much each vehicle is close to its deadline. One method to reduce the state space size is the concept of grouping vehicles\cite{walraven2016planning}. Vehicles are grouped based on their temporal proximity to their deadline. The freedom percentiles allocated in the state tuple serve as boundaries for grouping EVs.  Here, $Fr_{ave}$ is the average freedom of all EVs, and $Fr_x$ is the $x$ percentile of freedom. In this work, the features that lead to the best learning have been chosen by experiments and are presented as an 8-tuple.\\
The state of the system at each time is a vector constituted with scalars as $s_t=(t,n_{min},n_{max},T_t,Fr_{ave},Fr_2,Fr_5,Fr_{10})$. Here, $t$ is the time step of the day. The only decision variable in this problem is $a_t$, i.e. the number of EVs which are to be charged at each time step. In the solution methodology section, it is explained how $s_t$ and $a_t$ work together to obtain the Q-function. At each step, the actions of the system are done based on the priority of the EVs, which is obtained as given in \eqref{eq:priority}, and ensures that all $EV_i \in \mathcal{C}$ are charged at the next control step.
\begin{equation} \label{eq:priority}
    Pr_i = \begin{matrix}
sort(t_i^a-t_i^r) & i\in\mathcal{V}
\end{matrix} 
\end{equation}

One defined reward is such that the goal of the EVs is to follow a reference power which is known beforehand. The reference function could be any arbitrary curve. As the goal in this work is to demonstrate how EVs can be utilized to offer flexibility, the performance of EVs when set to follow a predetermined curve is discussed in the results section. The reward, in these cases, is defined as the negative of squared power mismatch between net charging power of EVs and the reference power, as in shown in \eqref{eq:case0}. Therefore, the reward will always be negative and agent tries to always reach smaller negative rewards.  It is also noteworthy that in the case studies, a problem setting is selected where the pre-purchased energy is roughly equal to the amount of daily energy that vehicles consume. That justifies the square usage in reward definition as it is crucial that EVs stay as close to the reference curve as possible at all times. 
\begin{equation}\label{eq:case0}
    r_t=-(P^{ref}_t - \sum_{i=1}^{N}p_{(i,t)})^2
\end{equation}
Another way to define the reward function is calculated as the amount of charging power consumption of the EV fleet which is coinciding with the PV generation. Based on \eqref{eq:case2_reward}, If any charging happens outside from the PV generation, no positive or negative reward is designated.
\begin{equation}\label{eq:case2_reward}
    r_{t}=\min\left\{PV_t, \sum_{i=1}^{N}p(i,t)\right\}
\end{equation}
It was mentioned that the state is an 8-tuple representation of the whole set of EVs. However, with non-deterministic PV outputs, measures need to be taken to improve state representation so the regression function can have better estimates of the stochastic nature of PV generations. To this end, two indicators $I^1_t$ and $I^2_d$ are introduced to the $s_t$. These features have to be chosen carefully in order to improve the learning experience and after several experiments, we have come up with $I^1_t$ as shown in \eqref{eq:pv indicator}.
\begin{equation} \label{eq:pv indicator}
    I^1_t = \sum_{k\in(t-60:\delta_c:t)}PV_k
\end{equation}
It is assumed that the power output of PV units is known for the past time steps of the day, and \eqref{eq:pv indicator} defines $I^1_t$ to be equal to the summation of PV generation over the past hour. $I^2_d$ is also a number that is assigned for each day and is obtained based on the predictions of the weather. It can have 3 values, with 3 for a clear day and 1 for a cloudy/rainy day with low solar irradiation. In the next section, a framework to solve the general environment settings with arbitrary behavior of EV owners, reference power and PV capacity is introduced.
\section {Solution Methodology}
As suggested earlier, fitted $Q$-iteration is chosen as the approach to solve the RL problem in this study. Basically, the optimal solution of an MDP is acquired by maximizing  the discounted sum of the rewards. The discount rate, $\gamma$, determines the present value of future rewards: a reward received $k$ time steps in the future is worth only $\gamma^{k-1}$ times what it would be worth if it were received immediately. At the time $t$, the value of taking action $a_t$ at state $s_t$ is denoted by $Q(s_t,a_t)$ and is calculated according to \textit{Bellman's optimality equation} as stated in \eqref{eq:qmax}.
\begin{equation}\label{eq:qmax}
    Q(s_t,a_t) = R(s_t,a_t) + \gamma \max_{a_{t+1}\in\mathcal{H}_{t+1}} Q(s_{t+1},a_{t+1})
\end{equation}
Based on \eqref{eq:qmax}, the action value function at each state is the sum of immediate reward at that state plus the maximum achievable action value at the subsequent state, for all feasible actions at time $t+1$. 
The procedure of $Q$-iteration used for solving the problem is presented in Algorithm \ref{alg_1}. This algorithm takes the discount factor ($\gamma$), the maximum number of simulation days (D), the maximum number of iterations ($K_{max}$), action and control step size ($\delta_{s}$ and $\delta_{c}$, respectively), the initial policy ($\pi_0$) and the target ($\epsilon_0$) for explorer as inputs; while returning the action value function as output. At the initializing step, the variables for day and time indices are set to zero and the sets containing batch samples for each day ($\mathcal{J}$) and the whole simulation period ($\mathcal{F}$) are pre-allocated.
\setlength{\textfloatsep}{0pt}
\begin{algorithm}[h!t] \label{alg_1}
    \SetKwInOut{Input}{Input}
    \SetKwInOut{Output}{Output}
    \SetKwInOut{Initiate}{Initiate}
    \caption{Fitted $Q$-Iteration Algorithm for EV Fleet Autonomous Charging Scheduling Problem}

    \Input{$\gamma, D, K_{max},\delta_s,\delta_c,\pi_0,\epsilon_0$}
    \Output{$Q_D^*(s_t,a_t)$}
    \Initiate{$\mathcal{F} \leftarrow \O, \mathcal{J} \leftarrow \O, t \leftarrow 0, d \leftarrow 0$}
    \While{$d < D$}
        {\textit{Initiate first day based on} $\pi_0$\\
        \For {$t=0 : \delta{s}:1440$}
            {
            \eIf {mod(t, $\delta c$) = 0}
                {
                    $a^*\leftarrow {arg}\max_{a_{t}\in\mathcal{H}_t} Q_d^*(s_{t},a_{t})$\\
                    $a_t=Explorer(a^*,\mathcal{H}_t,\epsilon)$\\
                    $s_{t+1},r_t\leftarrow env(s_t,a_t)$\\
                    $\mathcal{J}=\mathcal{J}\cup(s_{t-\delta{s}},s_t,r_t,a_t,\mathcal{H}_t)$
                }
                {
                $a_t\leftarrow a_{(t-\delta{s})}$\\
                $s_{t+1},r_t\leftarrow env(s_t,a_t)$\;
                }
            }
        $\mathcal{F} \leftarrow \mathcal{F} \cup \mathcal{J}$\\
        $d\leftarrow {d+1}$\\
        $Q_{d+1}^0(s_{t},a_{t})\leftarrow \O$\\
        \For{$k=1 : K_{max}$}
            {
            \For{$l=1 : |\mathcal{F}|$}
                {
                $X^l\leftarrow (s_t,a_t)$\\
                $Y^l\leftarrow r_t + \gamma \max_{a_{t+1}\in\mathcal{H}_{t+1}} Q(s_{t+1},a_{t+1})$\\
                }
            $Q_{d+1}^{k+1}(s_{t},a_{t})\leftarrow Regress(X^l,Y^l)$
            }
        $Q_{d+1}^*(s_{t},a_{t})\leftarrow Q_{d+1}^{K_{max}}(s_{t},a_{t})$\\
    }
\end{algorithm}
\\Line 2 of Algorithm \ref{alg_1} asserts that for the initial day ($d=0$), the actions are acquired based on the initial policy, i.e. $\pi_0(s_t)$. The preset policy is used only for the first day, and we have implemented the random policy for this purpose where actions at each state are chosen randomly from the feasible set of actions, $\mathcal{H}_t$. For the rest of the days, the actions are executed at time steps for each $\delta_s$ minutes; however, it is only at control time steps where $mod(t,\delta_c)=0$ that new actions are chosen. According to line 5, the best action is first obtained from the optimal action value function. Then based on line 6, the action at each time step is chosen by the explorer. To this end, we have implemented the $\epsilon$-greedy exploration method, where optimal action is acquired as shown in \eqref{eq:e-greedy1} and \eqref{eq:e-greedy2}.

\begin{equation}\label{eq:e-greedy1}
    \epsilon=\max\left\{\epsilon_0, 1-c_0d\right\}
\end{equation}
\begin{equation}\label{eq:e-greedy2}
    a_t = \left\{\begin{matrix}
a^*, & \epsilon\leq\rho\\
&\\
random(\mathcal{H}_t), & \epsilon>\rho
\end{matrix}\right. \end{equation}

A linear $\epsilon$-greedy method is selected here, where according to \eqref{eq:e-greedy1}, the $\epsilon$ curve is a line decreasing from 1 to $\epsilon_0$ with a slope equal to $c_0$. The action at each step is then extracted based on  \eqref{eq:e-greedy2}, where $\rho$ is a random number between 0 and 1. This equation means with the probability of $1-\epsilon$, action at each time step is set to the optimal action returned from the action value function, and with the probability of $\epsilon$, actions are chosen randomly from the feasible action set. In lines 7 and 8 of  Algorithm \ref{alg_1}, the chosen action is forwarded to the environment and the pair $(s_{t+1},r_t)$, representing next state and reward for taking action $a_t$ is returned. Then, the set of batch experiences is updated.\newline
At the end of each day, set $\mathcal{F}$, which  has the total simulation experiences so far, is updated. Then through an iterative procedure as indicated in lines 17-24, input and output data are fed to the regression function. Finally, the optimal action value function estimator which is used at line 8 for calculating the optimum action choice is set to be equal to the approximator function of the last iteration. To achieve an estimate of convergence, the factor convergence rate is introduced in \eqref{eq:convergence}. $C^k_d$ is the learning convergence for day $d$ at iteration $k$, and is a measure to indicate how best action-value function at each iteration performs compared to the previous iteration. The mean of best action value is denoted by $\overline{Q}$.
\begin{equation}\label{eq:convergence}
C^k_d = (\overline{Q}^{k+1}_d-\overline{Q}^{k}_d)^2/{\overline{Q}^{k}_d}^2
\end{equation}
 The most important element of the fitted $Q$-iteration algorithm is the regression function. At the end of each epoch, the data from environment is fed to the regression function to acquire an approximation of the true Q values. Also, the agent depends on the regression function for choosing actions at each step. Some regression methods that are used by different authors include but are not limited to neural network approximators, multi layer perceptrons, support vector machines, decision trees, and random forests. The fitted $Q$-iteration algorithm requires fitting of any arbitrary (parametric or non-parametric) approximation architecture to the Q-function with no bias on the regression function.\\In this work, we follow the path of the original fitted $Q$-iteration algorithm, which made use of an ensemble of decision trees. \cite{ernst2005tree}. In the search for a desirable regression algorithm which is able to model any Q-function,  tree-based models are found to offer great flexibility, meaning they can be used for predicting any type of Q-function. They are non-parametric, i.e. they do not need repeated occasions of trial and errors for tuning the parameters. Tree based models are also computationally efficient and despite some other methods, with the increase in problem dimensions, the computation burden of tree based methods does not increase exponentially. Among tree based models, we opted for random forests. Random forest are created by putting together various decision trees. At each step, random forest algorithm selects a random subset of features. That is why this method is robust to outliers. The convergence of tree based regression for fitted $Q$-iteration is investigated in \cite{ernst2005tree}. All in all, they are a great tool for estimating a priori unknown shaped Q-functions with satisfactory performance.

\section {Illustrative Example}
In this section, several case studies are presented to illustrate the merit of the presented approach. Various settings are simulated and results for each case are discussed.  The parameters used for simulations are as follows: $\gamma$ = 0.95, $K_{max}$ = 25, $\delta_s$ = 10 (min), $\delta_c$ = 20 (min), $\epsilon_0$ = 0.05, and $c_0$ = 2.5. Total number of 100 EVs are considered in the simulations, with the battery ratings of 5$KW$ maximum power and 16.67$KWh$ capacity.
 
\subsection{Case 0 - Aggregator Load Following}
\begin{figure}[b!]
\centering
{\includegraphics[width=\linewidth]
{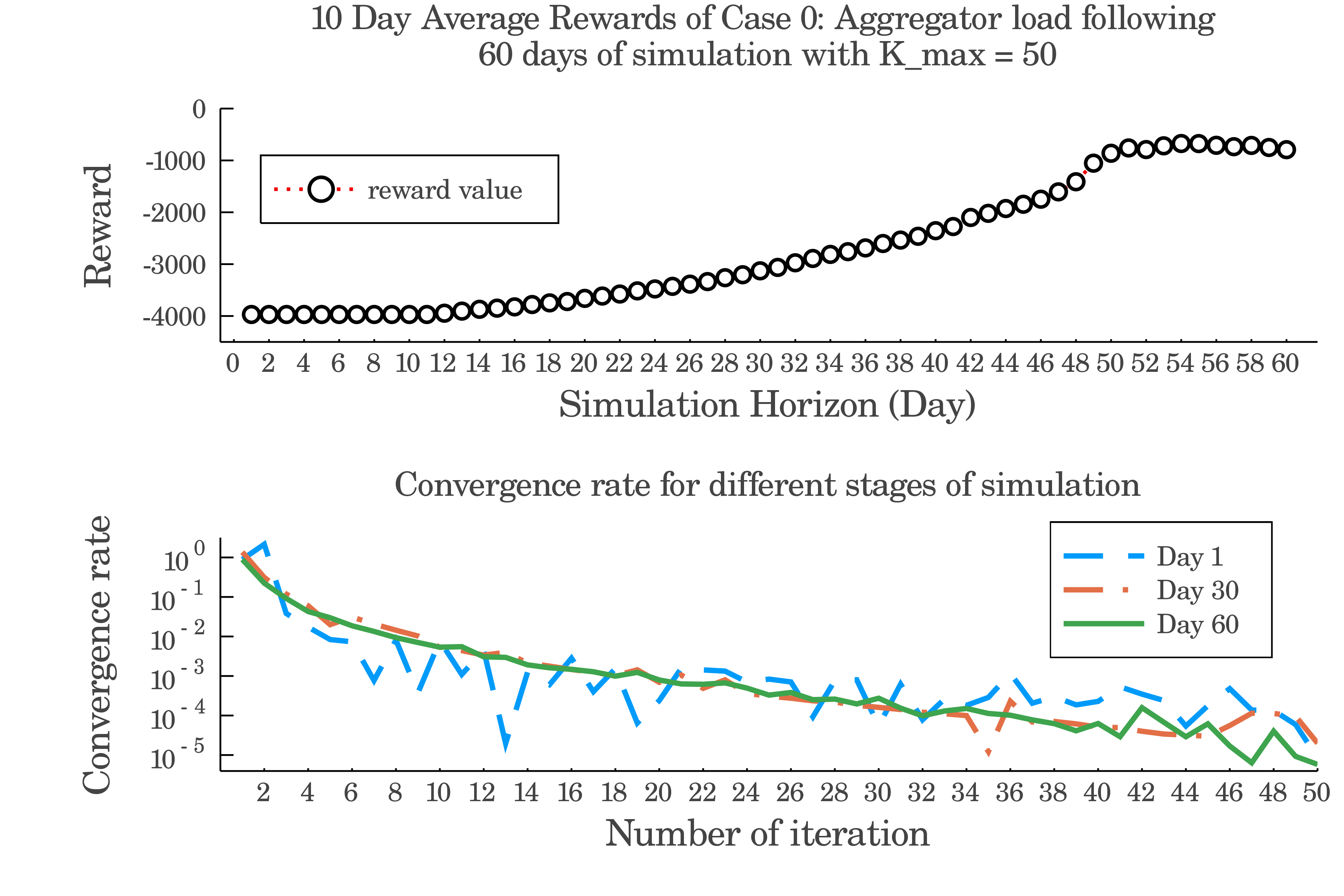}}
\caption{The Reward and Convergence in Case 0 for $D=60$ and $K_{max}=50$}
\label{fig:case0_reward}
\vspace{0.35cm}
{\includegraphics[width=\linewidth]
{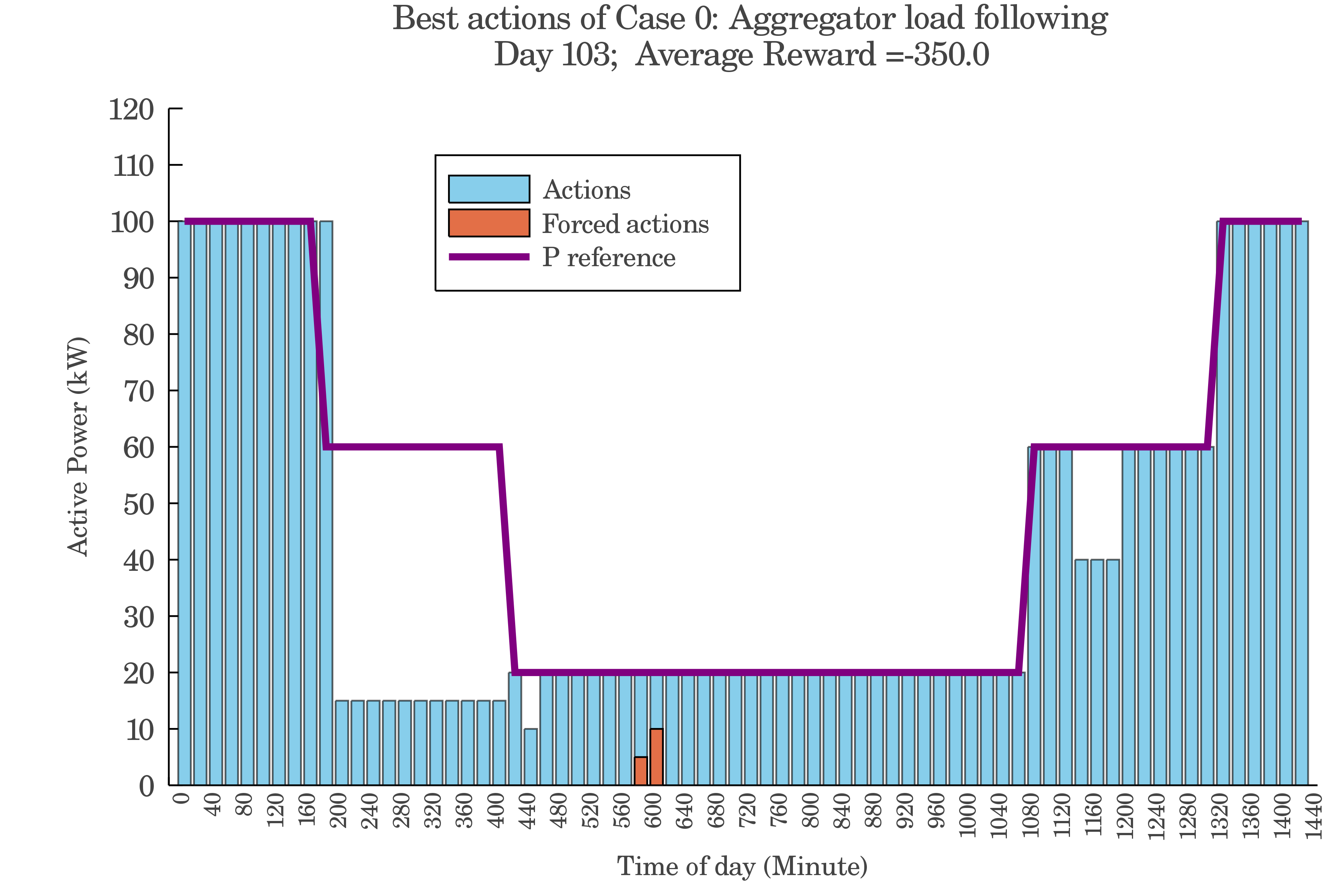}}
\caption{The Charging Power Curve in Case 0 for Day 103}
\label{fig:case0_day103}
\vspace{0.35cm}
{\includegraphics[width=\linewidth]
{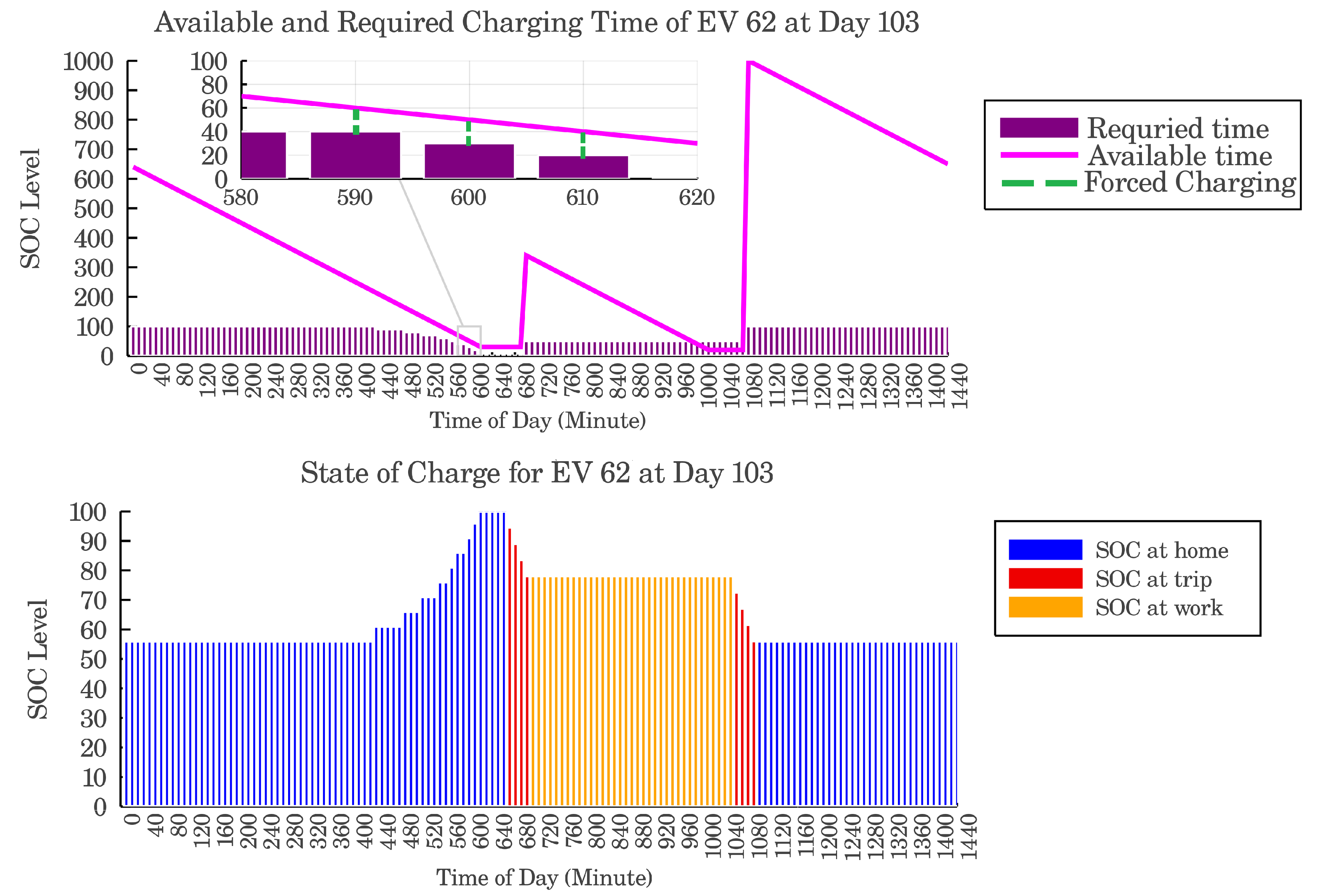}}
\caption{The Specifications of $EV_{62}$ in Case 0 for Day 103}
\label{fig:case0_ev62_d103}
\end{figure}

In this case, it is assumed that an aggregator already bought a specified amount of daily energy (purchased power agreement), and EV fleet is used to match the reference power with its charging behavior.  For this case, $D$ is considered to be equal to 75, and reward function \eqref{eq:case0} is used.  Fig. \ref{fig:case0_reward} shows the daily reward and convergence rate during learning process in this case. It is noticed that the reward function is increasing over time and stabilizes after 60 days. Also, based on the convergence rate, the error rate is decreased to as low as $10^{-5}$ which means 50 iterations are enough to reach acceptable approximations. As a result, $Q^*$ is obtained which is used to produce the best actions of the next days.
\\\color{black}In Case 0, the learning process lasts for 75 days. The results for applying the obtained Q-function to decide best actions on day 103 are displayed in Fig. \ref{fig:case0_day103}. In this figure, ``Actions" bars represent the collective amount of charging power for all EVs. It is observed in Fig. \ref{fig:case0_day103} that the total charging power is equal to the reference power throughout most of the day. It is also noticed that at minutes 580 and 600 of day 103, some actions are forced. To take a closer look at what happens during these periods, Fig. \ref{fig:case0_ev62_d103} is presented. In this figure, the daily required and available charging time, as well as SOC changes for $EV_{62}$ are illustrated in the top and bottom figures, respectively. As stated in \eqref{eq:set critical}, EVs that their required charging time is within the $\delta_c$ minutes of their available charging time, are forced to charge and $EV_{62}$ is one of them on day 103. The top figure is zoomed in during minutes 580-620 of the day, where the purple filled bars represent the amount of required charging time. At minute 590 of day 103, $t_{(a,62)}-t_{(r,62)}=20\leq\delta_c$ and thus, $EV_{62}$ is put in the set of critical EVs and is forced to charge at this time step. It is observed that this EV loses 45\% of its SOC as a result of daily communications and at 200 minutes before its morning departure, its SOC is only at 60\% which increases its risk of being placed in the critical set.\color{black}
\subsection{Case 1 - Ramp Following}
\begin{figure}[ht!]
\centering
{\includegraphics[width=\linewidth]
{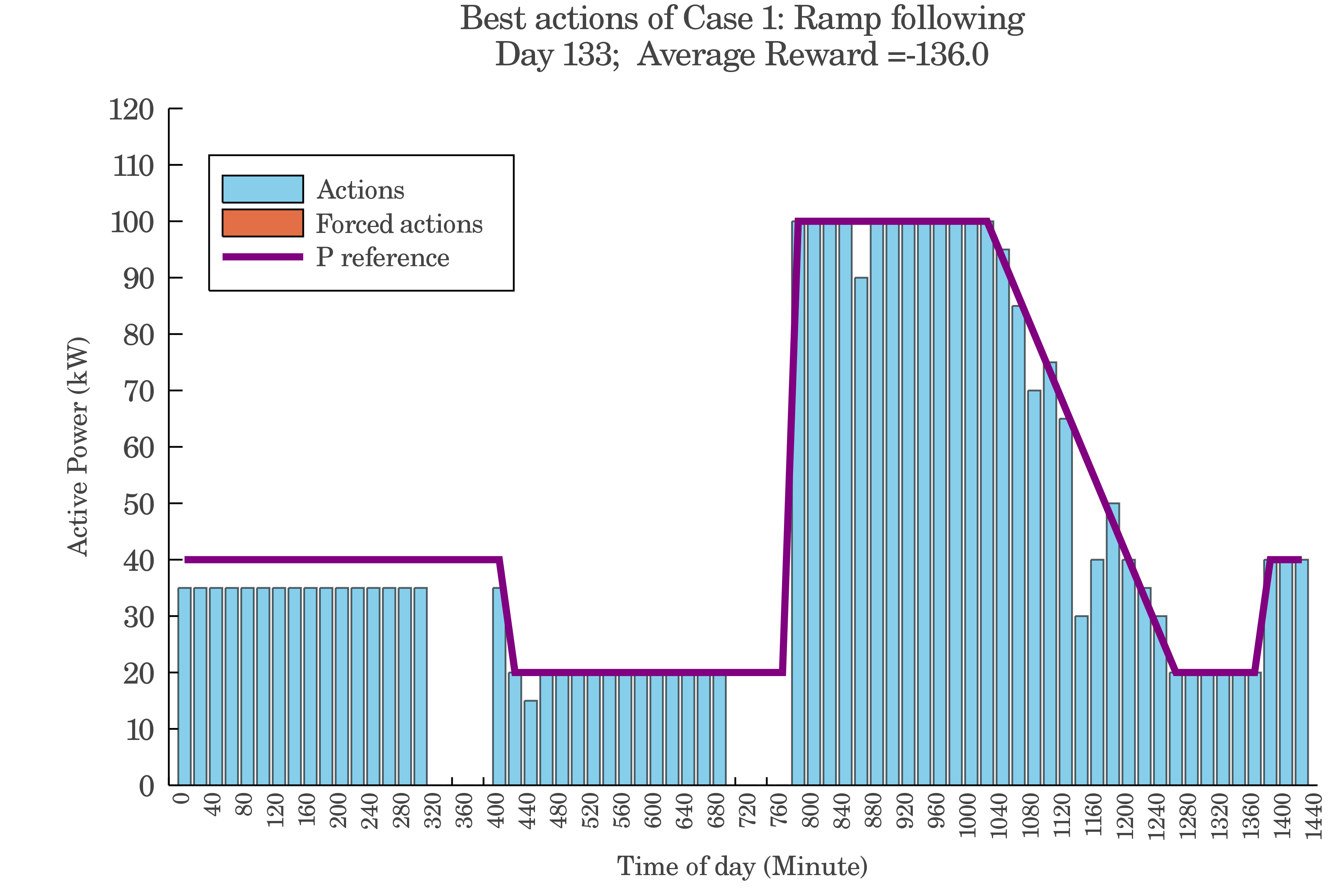}}
\caption{The Charging Power Curve in Case 1 for Day 133}
\label{fig:case1_day133}
\vspace{0.35cm}
{\includegraphics[width=\linewidth]
{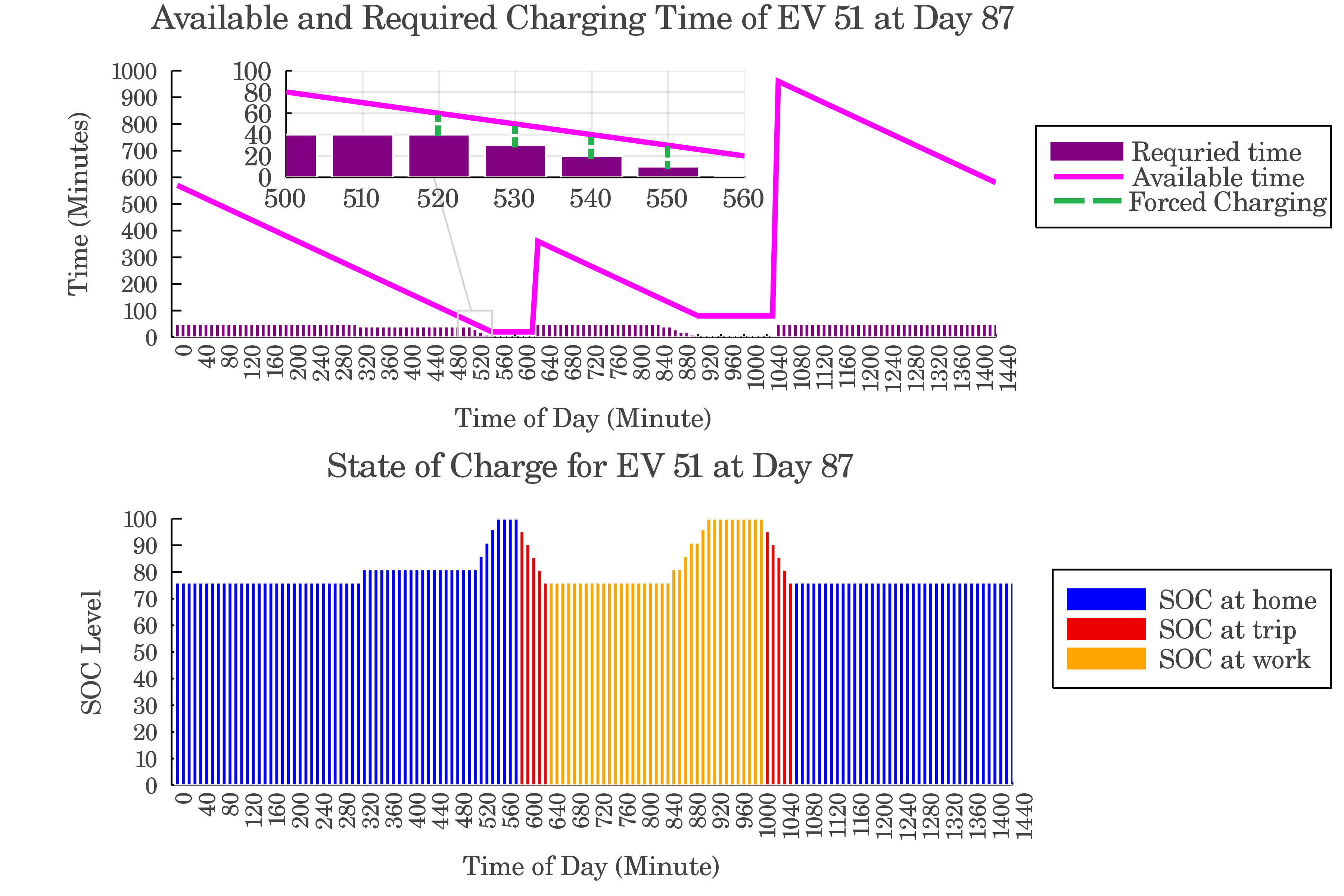}}
\caption{The Specifications of $EV_{51}$ in Case 1 for Day 87}
\label{fig:case1_ev51_d87}
\end{figure}

In this case, it is assumed that the EVs belong to an aggregator who uses them to offer ramping services. In a demand response service, EVs offer to lower their consumption during ramping hours of the system. This time, the reference power curve is according to Fig. \ref{fig:case1_day133}, with a steep reduction during the 4-hour period of 5 $p.m$ to 9 $p.m$.  For this case, $D$ is considered to be equal to 75, and reward function \eqref{eq:case0} is used.  It is seen that EVs learned effectively to alter their charging patterns according to the reward signal, which is defined the same as in Case 0. 
For day 87, the charging curve of $EV_{51}$ is shown in Fig. \ref{fig:case1_ev51_d87}. 
Through minutes 320 to 520 of day 87, a lot of charging potential is wasted and many EVs are forced to charge after minute 520 to reach full battery SOC before their departure, $EV_{51}$ being one of them. Although according to Fig. \ref{fig:case1_ev51_d87}, this EV is fully charged during working hours and its required charging is reduced to half, it is placed among critical units at minutes 520 through 550 due to be postponing its charging.
\subsection{Case 2 - Enhancing PV generation Dispatchability}
\begin{figure}[h!]
\centering
{\includegraphics[width=\linewidth]
{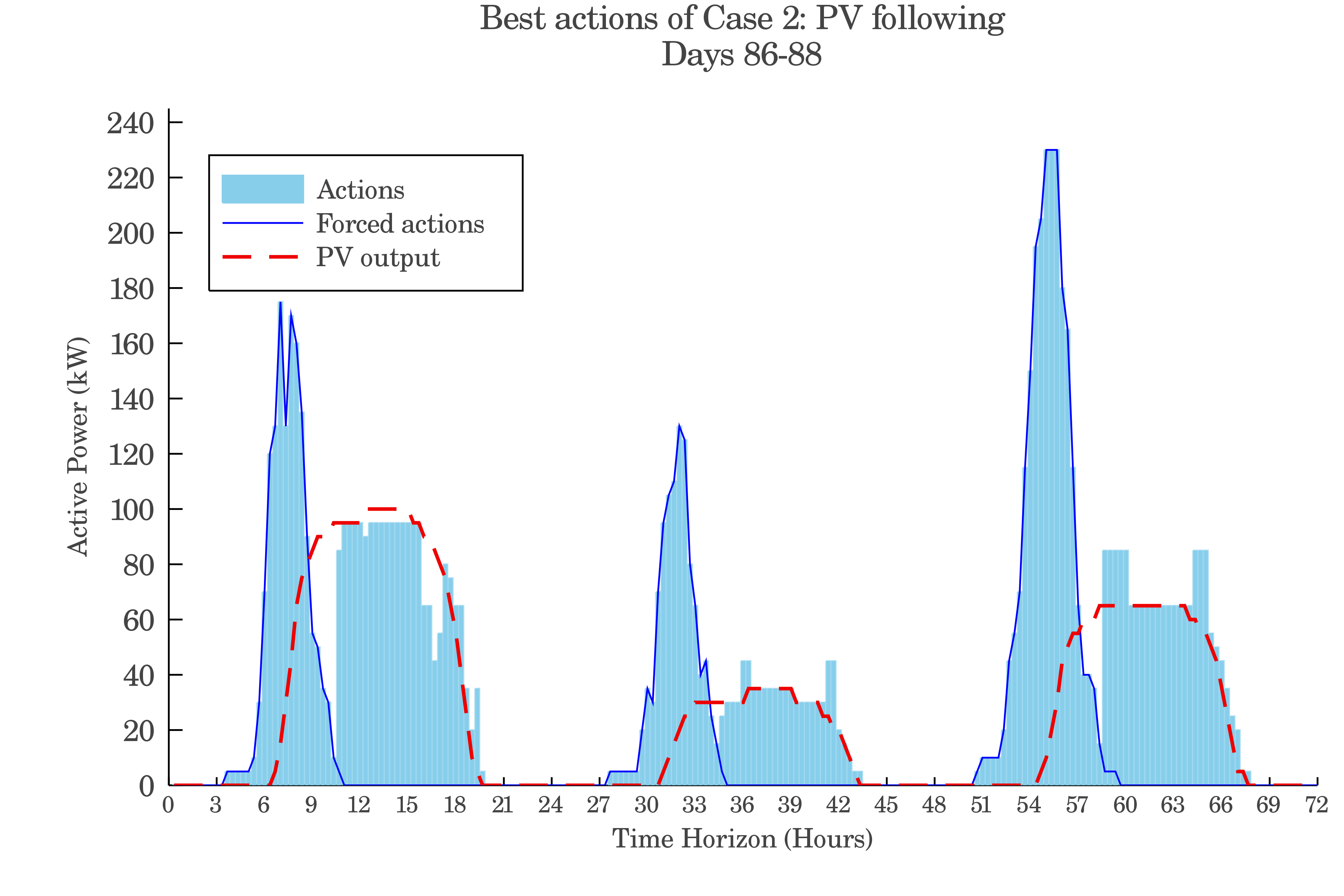}}
\caption{The Charging Power Curve in Case 2 for Days 86-88}
\label{fig:case2_3days}
\end{figure}

Here, it is assumed that EVs belong to an aggregator who also owns some PV units and their objective is to consume as much during PV generation as possible to minimize the unused PV capacity.  For this case, $D$ is considered to be equal to 60, and reward function \eqref{eq:case2_reward} is used. 
The results of this case suggest that in this way, the EVs offer a good solution to reduce PV curtailments, especially for days where PV output is higher and the most curtailments happen. Fig. \ref{fig:case2_3days} shows the result of a 3-day scheduling horizon for EV fleet in case 2. 

\begin{figure}[h!]
\centering
{\includegraphics[width=\linewidth]
{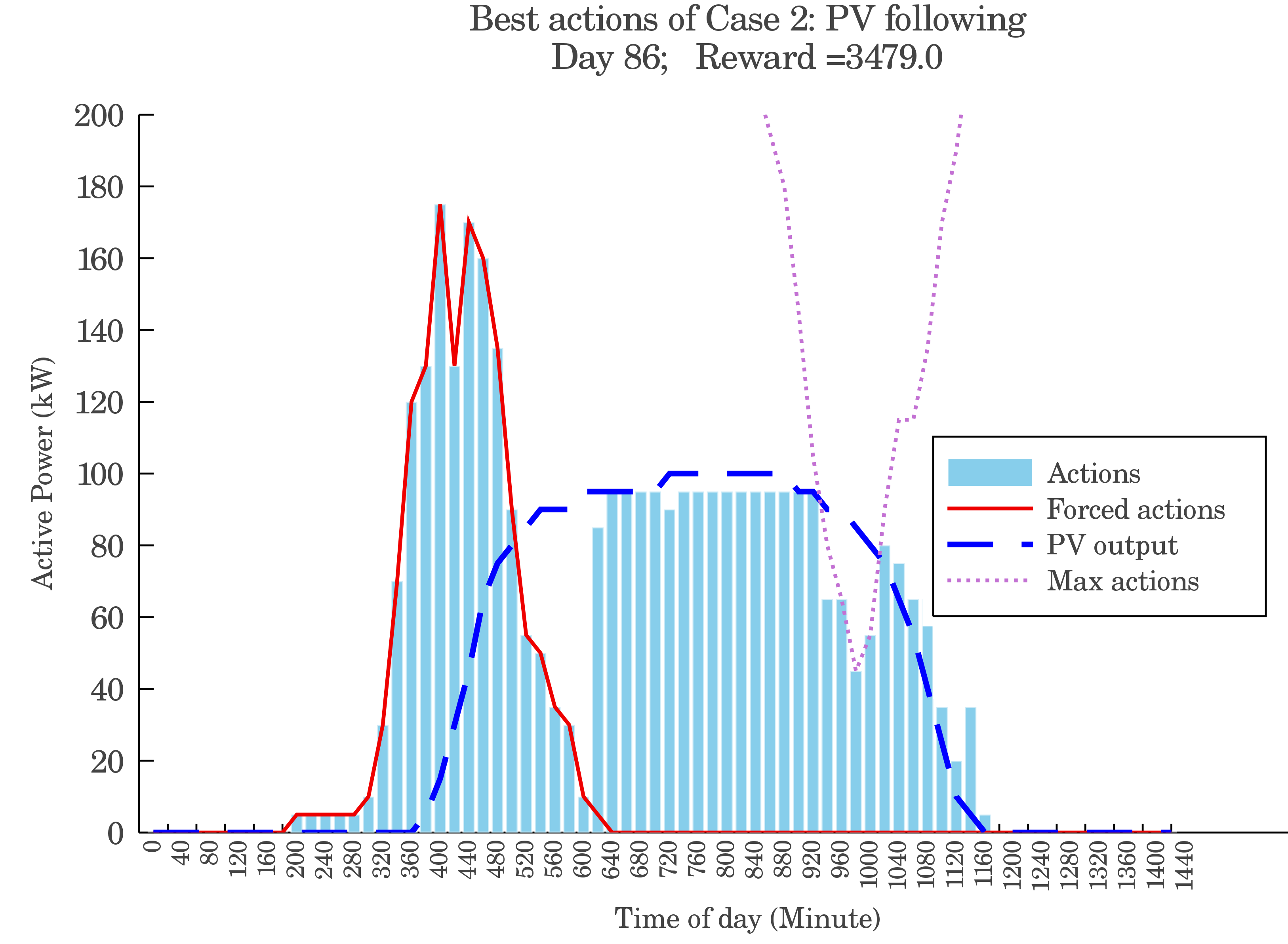}}
\caption{The Charging Power Curve in Case 2 for Day 86}
\label{fig:case2_1day}
\end{figure}

It is seen from Fig. \ref{fig:case2_3days} that for a random 3 day running horizon, PV is exploited to a very good extend. The solid black line shows the forced actions, i.e. instances where forced charging has happened. \color{black} It is also observed in Fig. 8 that the amount of forced actions in each day are related to the amount of PV generation in the previous day. The more PV generation happens during a day, the less forced actions occur at the next day. EVs have learned to charge with solar output, so if more PV generation is available, more energy is charged. As EVs are not rewarded for charging outside of PV generation hours, they show no inclination towards being charged during those hours and take no charging action until ``forced charging" takes place. Because EVs have stored more energy during day 86 than day 87, the forced actions in the beginning of day 87 are less than day 88.\color{black}\\
Fig. \ref{fig:case2_1day} illustrates the charging patterns happening during the first day in the three day horizon displayed in Fig. \ref{fig:case2_3days}. It is observed that all of the actions done before PV generation takes place are forced. This means that EVs have learned to delay their charging as much as possible so they can be charged when PVs start to generate. Also, it is noted that during the minutes 940 through 1000 on this day, the maximum possible actions are lower than the PV outputs. It means at this period, no further EVs can be charged to meet the PV generation fully.

\begin{figure}[h!]
\centering
{\includegraphics[width=\linewidth]
{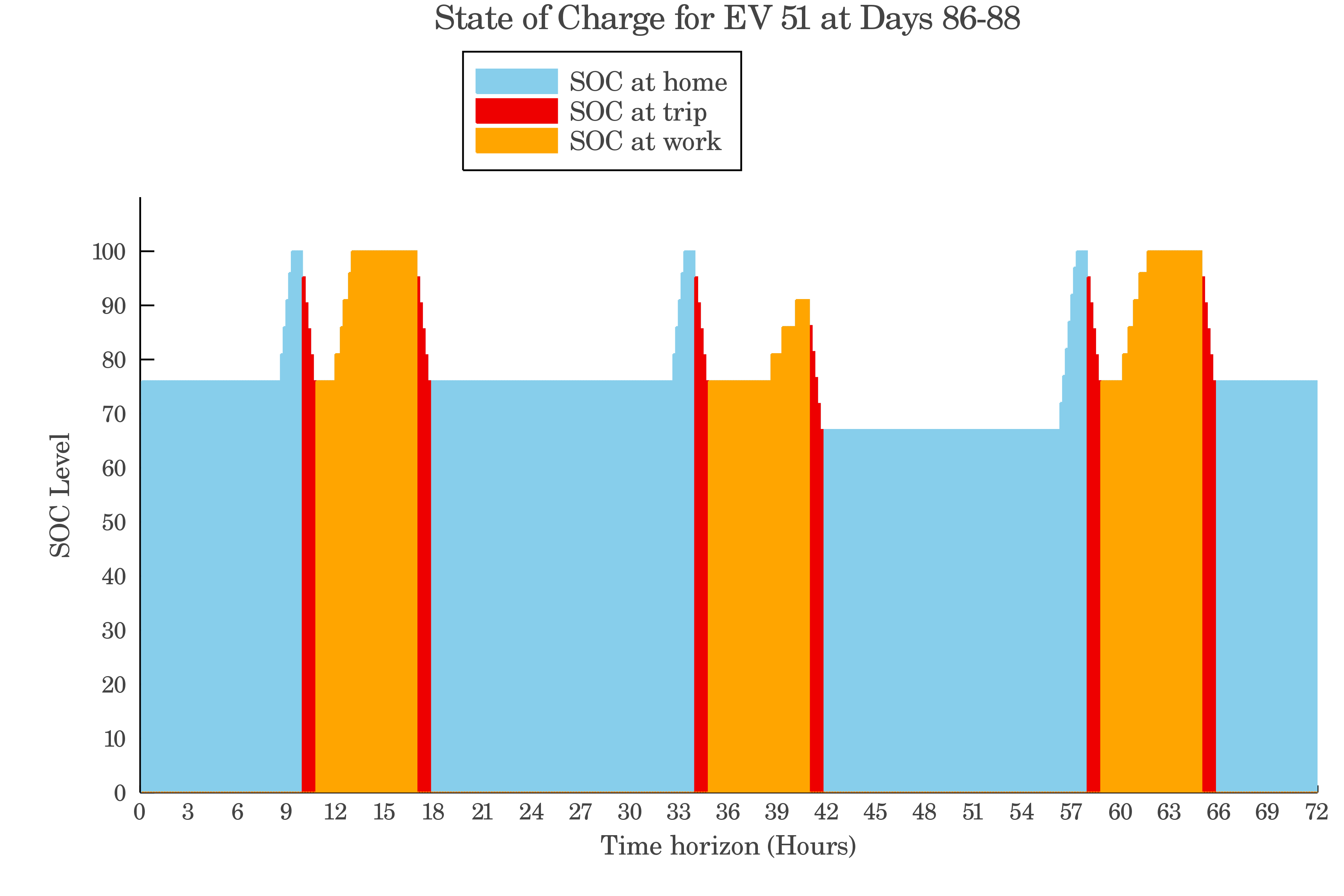}}
\caption{SOC of $EV_{51}$ in Case 2 for Days 86-88}
\label{fig:case2_soc51}
\end{figure}

Fig. \ref{fig:case2_soc51} shows the SOC trend of $EV_{51}$ for the three day horizon displayed in Fig. \ref{fig:case2_3days}, i.e. days 86 through 88. It is detected that at around hour 9 in the morning each day, this EV is set to depart and must be charged to 100\% SOC before that time. For the first and third days in which PV outputs are at higher levels, this EV is able to charge fully during its stay at the workplace charging station. For the second day, which PV outputs are the lowest, this EV learned to charge during the PV generation too; however, this time it can not reach 100\% SOC due to the lack of renewable generation.

The performance comparison of the learned autonomous behavior during different stages of learning with the optimal solution from all-knowing optimizer is summarized in table \ref{table-comparison}.  The all-knowing optimizer is supposed to know all of the stochastic para metres beforehand and returns the global optimum action at each decision step. The actions taken by the all-knowing oracle are acquired by leveraging a linear programming problem formulation.  The results at the start of training with no prior experience are brought as initial phase. The results after 36 days of training and final stage of training are also shown. It is observed that by moving ahead in time, the performance improves substantially and with experiencing more days during simulation, the performance of the learned Q-function will be better. 
\begin{table}[h!]
	\small \centering
	\caption{  performance comparison of the learned autonomous behavior} 

	\begin{tabular}{llll} \hline 
\textbf{Case Study} & \textbf{Case0} & \textbf{Case1} & \textbf{Case2} \\\hline
Performance Measure & Reward & Reward  & PV Utilization \\ 
Learning Days  & 75 & 75 &  60 \\ \hline\hline
\textbf{Benchmark Optimizer}  & -105 & -30 & 100\% \\ \hline
\textbf{Initial Phase}  & -5468 & -5333 & 58.3\% \\ \hline
\textbf{After 36 Days}  & -4201 & -3646 &  78.1\%  \\ \hline
\textbf{Final Performance}  & -350 & -136 & 90.1\% \\ \hline\hline
	\end{tabular}
	\label{table-comparison}
\end{table}

\subsection{Fitted Q-Iteration vs. DRL}
\begin{figure}[h!]
\centering
{\includegraphics[width=\linewidth]
{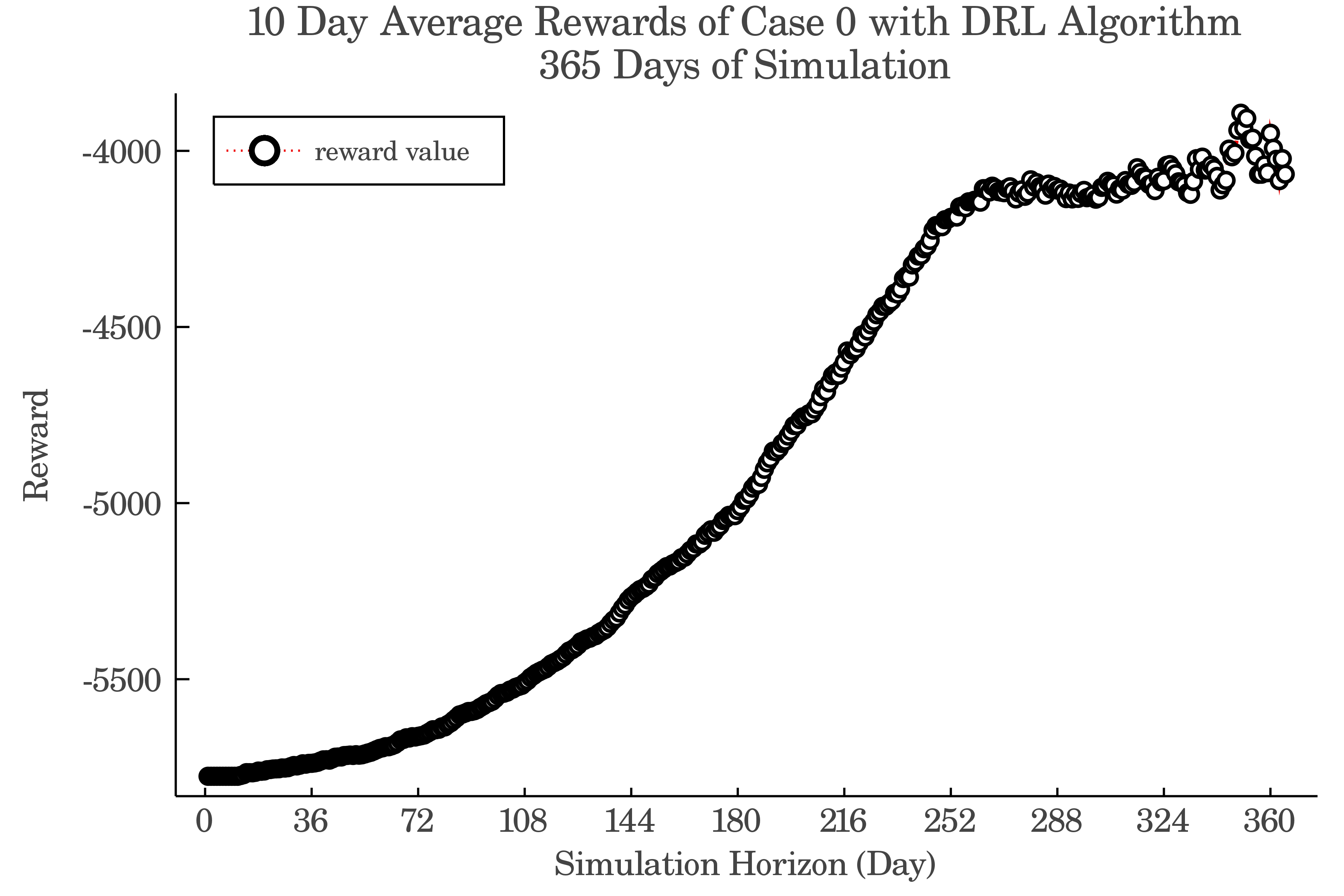}}
\caption{ The Average Rewards of DRL algorithm in Case 0 for 1 Year Simulation}
\label{fig:drl_reward}
\end{figure}

In this section, the results of deploying a DRL algorithm for the EV scheduling problem are displayed. The algorithm makes use of deep neural networks to reach an approximation of the Q value an is similar to the one used in \cite{wan2018model}. The results for learning the problem in Case 0 are illustrated and compared with the results of employing fitted $Q$-iteration for the same case. It is shown the the DRL approach needs substantially more data to reach satisfactory reward values. \\
The rewards curve during the learning process with DRL algorithm is presented in Fig \ref{fig:drl_reward}. The rewards for the first 75 days of simulation in Case 0 for fitted $Q$-iteration and DRL algorithm are compared with each other in Table \ref{table-drl}. The rewards of three stages of learning are presented in the columns of this table. It is observed from Table \ref{table-drl} that after 75 days of learning, the DRL method's reward is still very low, compared to that of fitted $Q$-iteration. According to the learning curve of DRL algorithm for 365 days of simulation which is plotted in Fig. \ref{fig:drl_reward}, the rewards are not still reaching competent values. These observations suggest that the neural network's overall learning rate is slower than the fitted $Q$-iteration algorithm for this problem, i.e. it needs a larger amount of training data to reach acceptable levels of Q approximation.

\begin{table}[h!]
	\small \centering
	\caption{  learning comparison of fitted $Q$-iteration vs. DRL} 
	\begin{tabular}{lll} \hline 
\textbf{Algorithm} & \textbf{Fitted $Q$-Iteration} & \textbf{DRL} \\\hline
Performance Measure & Reward & Reward \\ \hline\hline
\textbf{Benchmark Optimizer}  & -105 & -105 \\ \hline
\textbf{Initial Phase}  & -5468 & -6568 \\ \hline
\textbf{After 36 Days}  & -4201 & -6243 \\ \hline
\textbf{After 75 Days}  & -350 & -5732 \\ \hline\hline
	\end{tabular}
	\label{table-drl}
\end{table}

\section{Conclusion}
In this paper, an EV fleet was utilized to offer system flexibility as a means of offering a solution for reducing renewable curtailments. Three case studies were simulated which their results suggested that with appropriate training strategy, a set of EV vehicles can be implemented successfully for flexibility provision and curtailment reduction. The reinforcement algorithm used to solve the problem is fitted $Q$-iteration with the decision trees as the regression function. With this approach, instead of solving an optimization problem for each period, we simply look up the best action value from the Q-function.\\
In case 0, EVs learn to follow a reference power trajectory which is according to the amount of energy bought for the aggregator who owns parking stations. In case 1, it is shown that the EV fleet can also be offered as a resource for ramping products, in this case ramping down their consumption. Finally, for a system with integrated PV generations, in case 2 it is observed that EVs can learn how to delay their charging in order to utilize the most PV generation possible, and eventually help reduce the curtailments.  It is also shown that DRL algorithm can not reach a competent learning behavior with small amount of data in the EV scheduling problem of this article.  Currently, up to 18\% of PV generations in some days at CAISO are curtailed, and by coordination of EVs charging, our study showed great promise in this approach which has never been studied before. \newline
For future work, some possibilities to further explore this problem include deploying a multi agent framework for this problem, and searching for methods to cope with more complex uncertainty scenarios.

\bibliographystyle{IEEEtran}
\bibliography{second_jrnl.bbl}

\begin{thebibliography}{10}
\providecommand{\url}[1]{#1}
\csname url@samestyle\endcsname
\providecommand{\newblock}{\relax}
\providecommand{\bibinfo}[2]{#2}
\providecommand{\BIBentrySTDinterwordspacing}{\spaceskip=0pt\relax}
\providecommand{\BIBentryALTinterwordstretchfactor}{4}
\providecommand{\BIBentryALTinterwordspacing}{\spaceskip=\fontdimen2\font plus
\BIBentryALTinterwordstretchfactor\fontdimen3\font minus
  \fontdimen4\font\relax}
\providecommand{\BIBforeignlanguage}[2]{{%
\expandafter\ifx\csname l@#1\endcsname\relax
\typeout{** WARNING: IEEEtran.bst: No hyphenation pattern has been}%
\typeout{** loaded for the language `#1'. Using the pattern for}%
\typeout{** the default language instead.}%
\else
\language=\csname l@#1\endcsname
\fi
#2}}
\providecommand{\BIBdecl}{\relax}
\BIBdecl

\bibitem{eia2020}
E.~I. Administration, ``Annual energy outlook 2020 with projections to 2050,''
  2020, Accessed: May 5, 2020. [Online]. Available:
  https://www.eia.gov/outlooks/aeo/.

\bibitem{kikusato2019electric}
H.~Kikusato, Y.~Fujimoto, S.-i. Hanada, D.~Isogawa, S.~Yoshizawa, H.~Ohashi,
  and Y.~Hayashi, ``Electric vehicle charging management using auction
  mechanism for reducing pv curtailment in distribution systems,'' \emph{IEEE
  Transactions on Sustainable Energy}, 2019.

\bibitem{golden2015curtailment}
R.~Golden and B.~Paulos, ``Curtailment of renewable energy in california and
  beyond,'' \emph{The Electricity Journal}, vol.~28, no.~6, pp. 36--50, 2015.

\bibitem{caisowebsite}
CAISO, ``California independent system operator website,'' 2020, Accessed: May
  5, 2020. [Online]. Available: https://www.caiso.com/.

\bibitem{cochran2015grid}
J.~Cochran, P.~Denholm, B.~Speer, and M.~Miller, ``Grid integration and the
  carrying capacity of the us grid to incorporate variable renewable energy,''
  National Renewable Energy Lab.(NREL), Golden, CO (United States), Tech. Rep.,
  2015.

\bibitem{cochran2012integrating}
J.~Cochran, L.~Bird, J.~Heeter, and D.~J. Arent, ``Integrating variable
  renewable energy in electric power markets. best practices from international
  experience,'' National Renewable Energy Lab.(NREL), Golden, CO (United
  States), Tech. Rep., 2012.

\bibitem{bird2014wind}
L.~Bird, J.~Cochran, and X.~Wang, ``Wind and solar energy curtailment:
  Experience and practices in the united states,'' National Renewable Energy
  Lab.(NREL), Golden, CO (United States), Tech. Rep., 2014.

\bibitem{li2015comprehensive}
C.~Li, H.~Shi, Y.~Cao, J.~Wang, Y.~Kuang, Y.~Tan, and J.~Wei, ``Comprehensive
  review of renewable energy curtailment and avoidance: a specific example in
  china,'' \emph{Renewable and Sustainable Energy Reviews}, vol.~41, pp.
  1067--1079, 2015.

\bibitem{henriot2015economic}
A.~Henriot, ``Economic curtailment of intermittent renewable energy sources,''
  \emph{Energy Economics}, vol.~49, pp. 370--379, 2015.

\bibitem{ela2009using}
E.~Ela, ``Using economics to determine the efficient curtailment of wind
  energy,'' National Renewable Energy Lab.(NREL), Golden, CO (United States),
  Tech. Rep., 2009.

\bibitem{zou2015mitigation}
J.~Zou, S.~Rahman, and X.~Lai, ``Mitigation of wind output curtailment by
  coordinating with pumped storage and increasing transmission capacity,'' in
  \emph{2015 IEEE Power \& Energy Society General Meeting}.\hskip 1em plus
  0.5em minus 0.4em\relax IEEE, 2015, pp. 1--5.

\bibitem{cleary2015assessing}
B.~Cleary, A.~Duffy, A.~OConnor, M.~Conlon, and V.~Fthenakis, ``Assessing the
  economic benefits of compressed air energy storage for mitigating wind
  curtailment,'' \emph{IEEE Transactions on Sustainable Energy}, vol.~6, no.~3,
  pp. 1021--1028, 2015.

\bibitem{moradzadeh2014congestion}
M.~Moradzadeh, B.~Zwaenepoel, J.~Van~de Vyver, and L.~Vandevelde,
  ``Congestion-induced wind curtailment mitigation using energy storage,'' in
  \emph{2014 IEEE International Energy Conference (ENERGYCON)}.\hskip 1em plus
  0.5em minus 0.4em\relax IEEE, 2014, pp. 572--576.

\bibitem{denholm2019timescales}
P.~Denholm and T.~Mai, ``Timescales of energy storage needed for reducing
  renewable energy curtailment,'' \emph{Renewable energy}, vol. 130, pp.
  388--399, 2019.

\bibitem{hozouri2014use}
M.~A. Hozouri, A.~Abbaspour, M.~Fotuhi-Firuzabad, and M.~Moeini-Aghtaie, ``On
  the use of pumped storage for wind energy maximization in
  transmission-constrained power systems,'' \emph{IEEE Transactions on Power
  Systems}, vol.~30, no.~2, pp. 1017--1025, 2014.

\bibitem{zhang2018review}
D.~Zhang, X.~Han, and C.~Deng, ``Review on the research and practice of deep
  learning and reinforcement learning in smart grids,'' \emph{CSEE Journal of
  Power and Energy Systems}, vol.~4, no.~3, pp. 362--370, 2018.

\bibitem{mocanu2018line}
E.~Mocanu, D.~C. Mocanu, P.~H. Nguyen, A.~Liotta, M.~E. Webber, M.~Gibescu, and
  J.~G. Slootweg, ``On-line building energy optimization using deep
  reinforcement learning,'' \emph{IEEE transactions on smart grid}, vol.~10,
  no.~4, pp. 3698--3708, 2018.

\bibitem{szinai2020reduced}
J.~K. Szinai, C.~J. Sheppard, N.~Abhyankar, and A.~R. Gopal, ``Reduced grid
  operating costs and renewable energy curtailment with electric vehicle charge
  management,'' \emph{Energy Policy}, vol. 136, p. 111051, 2020.

\bibitem{chics2016reinforcement}
A.~Chi{\c{s}}, J.~Lund{\'e}n, and V.~Koivunen, ``Reinforcement learning-based
  plug-in electric vehicle charging with forecasted price,'' \emph{IEEE
  Transactions on Vehicular Technology}, vol.~66, no.~5, pp. 3674--3684, 2016.

\bibitem{chen2018indirect}
T.~Chen and W.~Su, ``Indirect customer-to-customer energy trading with
  reinforcement learning,'' \emph{IEEE Transactions on Smart Grid}, vol.~10,
  no.~4, pp. 4338--4348, 2018.

\bibitem{claessens2016convolutional}
B.~J. Claessens, P.~Vrancx, and F.~Ruelens, ``Convolutional neural networks for
  automatic state-time feature extraction in reinforcement learning applied to
  residential load control,'' \emph{IEEE Transactions on Smart Grid}, vol.~9,
  no.~4, pp. 3259--3269, 2016.

\bibitem{watkins1992q}
C.~J. Watkins and P.~Dayan, ``Q-learning,'' \emph{Machine learning}, vol.~8,
  no. 3-4, pp. 279--292, 1992.

\bibitem{ernst2005tree}
D.~Ernst, P.~Geurts, and L.~Wehenkel, ``Tree-based batch mode reinforcement
  learning,'' \emph{Journal of Machine Learning Research}, vol.~6, no. Apr, pp.
  503--556, 2005.

\bibitem{mnih2015human}
V.~Mnih, K.~Kavukcuoglu, D.~Silver, A.~A. Rusu, J.~Veness, M.~G. Bellemare,
  A.~Graves, M.~Riedmiller, A.~K. Fidjeland, G.~Ostrovski \emph{et~al.},
  ``Human-level control through deep reinforcement learning,'' \emph{Nature},
  vol. 518, no. 7540, pp. 529--533, 2015.

\bibitem{zhang2019deep}
Z.~Zhang, D.~Zhang, and R.~C. Qiu, ``Deep reinforcement learning for power
  system applications: An overview,'' \emph{CSEE Journal of Power and Energy
  Systems}, vol.~6, no.~1, pp. 213--225, 2019.

\bibitem{da2019coordination}
F.~L. Da~Silva, C.~E. Nishida, D.~M. Roijers, and A.~H.~R. Costa,
  ``Coordination of electric vehicle charging through multiagent reinforcement
  learning,'' \emph{IEEE Transactions on Smart Grid}, vol.~11, no.~3, pp.
  2347--2356, 2019.

\bibitem{qian2019deep}
T.~Qian, C.~Shao, X.~Wang, and M.~Shahidehpour, ``Deep reinforcement learning
  for ev charging navigation by coordinating smart grid and intelligent
  transportation system,'' \emph{IEEE Transactions on Smart Grid}, 2019.

\bibitem{ko2018mobility}
H.~Ko, S.~Pack, and V.~C. Leung, ``Mobility-aware vehicle-to-grid control
  algorithm in microgrids,'' \emph{IEEE Transactions on Intelligent
  Transportation Systems}, vol.~19, no.~7, pp. 2165--2174, 2018.

\bibitem{li2019constrained}
H.~Li, Z.~Wan, and H.~He, ``Constrained ev charging scheduling based on safe
  deep reinforcement learning,'' \emph{IEEE Transactions on Smart Grid}, 2019.

\bibitem{wan2018model}
Z.~Wan, H.~Li, H.~He, and D.~Prokhorov, ``Model-free real-time ev charging
  scheduling based on deep reinforcement learning,'' \emph{IEEE Transactions on
  Smart Grid}, vol.~10, no.~5, pp. 5246--5257, 2018.

\bibitem{shin2019cooperative}
M.~Shin, D.-H. Choi, and J.~Kim, ``Cooperative management for pv/ess-enabled
  electric-vehicle charging stations: A multi-agent deep reinforcement learning
  approach,'' \emph{IEEE Transactions on Industrial Informatics}, 2019.

\bibitem{wei2018electric}
Z.~Wei, Y.~Li, and L.~Cai, ``Electric vehicle charging scheme for a
  park-and-charge system considering battery degradation costs,'' \emph{IEEE
  Transactions on Intelligent Vehicles}, vol.~3, no.~3, pp. 361--373, 2018.

\bibitem{vandael2015reinforcement}
S.~Vandael, B.~Claessens, D.~Ernst, T.~Holvoet, and G.~Deconinck,
  ``Reinforcement learning of heuristic ev fleet charging in a day-ahead
  electricity market,'' \emph{IEEE Transactions on Smart Grid}, vol.~6, no.~4,
  pp. 1795--1805, 2015.

\bibitem{sadeghianpourhamami2019definition}
N.~Sadeghianpourhamami, J.~Deleu, and C.~Develder, ``Definition and evaluation
  of model-free coordination of electrical vehicle charging with reinforcement
  learning,'' \emph{IEEE Transactions on Smart Grid}, 2019.

\bibitem{mbuwir2017battery}
B.~V. Mbuwir, F.~Ruelens, F.~Spiessens, and G.~Deconinck, ``Battery energy
  management in a microgrid using batch reinforcement learning,''
  \emph{Energies}, vol.~10, no.~11, p. 1846, 2017.

\bibitem{manshadi2017wireless}
S.~D. Manshadi, M.~E. Khodayar, K.~Abdelghany, and H.~{\"U}ster, ``Wireless
  charging of electric vehicles in electricity and transportation networks,''
  \emph{IEEE Transactions on Smart Grid}, vol.~9, no.~5, pp. 4503--4512, 2017.

\bibitem{manshadi2019strategic}
S.~D. Manshadi and M.~E. Khodayar, ``Strategic behavior of in-motion wireless
  charging aggregators in the electricity and transportation networks,''
  \emph{IEEE Transactions on Vehicular Technology.}, 2020.

\bibitem{wang2020v2g}
X.~Wang, J.~Wang, and J.~Liu, ``V2g frequency regulation capacity optimal
  scheduling for battery swapping station using deep q-network,'' \emph{IEEE
  Transactions on Industrial Informatics}, 2020.

\bibitem{walraven2016planning}
E.~Walraven and M.~T. Spaan, ``Planning under uncertainty for aggregated
  electric vehicle charging with renewable energy supply,'' in
  \emph{Proceedings of the Twenty-second European Conference on Artificial
  Intelligence}, 2016, pp. 904--912.

\end{thebibliography}
\end{document}